\documentclass{aa}  
\usepackage{graphicx}
\usepackage{txfonts}

\usepackage{hyperref}

\usepackage[dvipsnames]{xcolor}
\usepackage{caption,subcaption} 

\usepackage{float}

\usepackage{multirow}
\usepackage{amsmath}
\usepackage{multicol,lipsum}

\begin{document}

   \title{Interior-atmosphere modelling to assess the observability of rocky planets with JWST}

   \author{L. Acuña \inst{1,2}
          \and M. Deleuil\inst{2,3}
          \and O. Mousis\inst{2,3}
          }

        \institute{Max-Planck-Institut für Astronomie, Königstuhl 17, D-69117 Heidelberg, Germany\\ \email{acuna@mpia.de}
        \and
        Aix-Marseille Université, CNRS, CNES, Institut Origines, LAM, Marseille, France
        \and
        Institut universitaire de France (IUF), France
        }

        \authorrunning{L. Acuña et al.}

   \date{Received 20 December 2022; accepted 24 April 2023}

 
  \abstract
   {Super-Earths present compositions dominated by refractory materials. However, there is a degeneracy in their interior structure between a planet that has no atmosphere and a small Fe content, and a planet that has a thin atmosphere and a higher core mass fraction. To break this degeneracy, atmospheric characterisation observations are required.}
   {We present a self-consistent interior-atmosphere model to constrain the volatile mass fraction, surface pressure, and temperature of rocky planets with water and CO$_{2}$ atmospheres. The parameters obtained in our analysis can be used to predict observations in emission spectroscopy and photometry with JWST, which can determine the presence of an atmosphere and, if present, its composition.}
   {We coupled a 1D interior model with a supercritical water layer to an atmospheric model. In order to obtain the bolometric emission and Bond albedo for an atmosphere in radiative-convective equilibrium, we used a low-resolution k-correlated atmospheric model within our retrieval of planetary mass, radius, and host stellar abundances. We generated emission spectra with the same atmospheric model at a higher resolution (R = 200-300). An adaptive Markov chain Monte Carlo was employed for an efficient sampling of the parameter space at low volatile mass fractions.}
   {From our interior structure retrieval, TRAPPIST-1 c is most likely to present a bare surface, although the presence of an atmosphere cannot be ruled out. We estimate a 1$\sigma$ confidence interval of the surface pressure for a water-dominated atmosphere of $P_{surf} = 40 \pm 40$ bar. We generated spectra for these two scenarios to compare with the emission flux of TRAPPIST-1 c recently observed in the MIRI F1500W filter. This is compatible with bare rock surfaces or a thin atmosphere with little or no CO$_{2}$.
   In the case of 55 Cancri e, a combined spectrum with NIRCam and MIRI LRS may present high uncertainties at wavelengths between 3 and 3.7 $\mu$m. However, this does not affect the identification of H$_{2}$O because it does not present spectral features in this wavelength range.
   }
   {}


   \maketitle
%

\section{Introduction}



Low-mass exoplanets ($M < 20 \ M_{\oplus}$) have two different sub-populations based on their radius and density: super-Earths and sub-Neptunes. Super-Earths have radii of $R = 1.3 \ R_{\oplus}$, while the radii of sub-Neptunes correspond to $R = 2.4 \ R_{\oplus}$ \citep{Fulton17,Fulton18}. If we compare these radii with planet interior and evolution models, super-Earths are mostly composed of Fe and Si-bearing rocks, whereas sub-Neptunes have a significant volatile (H/He, water) content. Despite having an idea of the main component for these planets from their mass and radius data and interior structure models, we do not know their exact interior composition due to degeneracies.

In the case of super-Earths, we still have the question of whether such a planet could have a thin atmosphere or a bare rock surface. Atmospheres containing H/He have been discarded since a very small fraction of H/He entails a minimum radius of $\simeq$ 1.6 $R_{\oplus}$ \citep{lopezfortney14}. Therefore, an atmosphere composed of water formed from ice pebbles accreted beyond or in the vicinity of the water ice line \citep{Mousis19,KrissansenTotton21,Kimura22}, a secondary atmosphere built\ up by outgassing \citep{Ortenzi20,Baumeister_platoconf,Liggins22}, or a silicate atmosphere \citep{zahnle10} are the most likely scenarios for super-Earths. This variety in the possible atmospheric composition produces a degeneracy in the internal structure of super-Earths and Earth-sized planets, as the same planetary mass and radius can be explained by a planet with no atmosphere and a low-Fe content rocky bulk \citep{Madhusudhan12,Dorn17a} or a planet with a thin atmosphere and a core mass fraction (CMF) similar to that of Earth (32\% of CMF).

This degeneracy in interior structure can only be broken with the support of atmospheric characterisation data. The presence of an atmosphere has been confirmed in the hot super-Earth $\pi$ Mensae c, whose detected  C II ions indicate atmospheric escape of a high molecular atmosphere \citep{Garcia21}. Phase curves have also been used to determine the existence of a silicate atmosphere in K2-141 b \citep{Zieba22}, and transmission spectroscopy has been used for the terrestrial planet LHS 3844 b \citep{Diamond20}. Moreover, \cite{Kreidberg19} use the phase curves to confirm the absence of an atmosphere, as well as to constrain which material constitutes the planetary surface. 

JWST \citep{jwst} will observe several super-Earths to confirm the presence of an atmosphere or even narrow their possible atmospheric compositions. In this study, we present a self-consistent interior-atmosphere model, Marseille's Super-Earth Interior model (MSEI), to perform retrievals from estimated mass, radius, and stellar host abundances. As a result, we determine the posterior distribution functions (PDFs) of the atmospheric mass as well as the surface pressure and temperature of water- and CO$_{2}$-dominated atmospheres. These atmospheric parameters obtained from our retrieval analysis can be used as input for an atmospheric model to produce spectra. We set an example of this application with our k-correlated atmospheric model, MSEIRADTRAN, to generate emission spectra to predict observations with JWST with the Mid-Infrared Instrument (MIRI) photometric filters and NIRCam and MIRI Low-Resolution Spectrometer (MIRI LRS). Combined interior and atmospheric models have been used to constrain the water mass fractions (WMF) of rocky planets as WMF < $10^{-3}$ \citep{Agol21}. However, the consistent exploration of the parameter space in the region close to WMF = 0 is necessary to accurately obtain the PDFs of the water mass fraction and the surface pressure. For this reason, we employ an adaptive Markov chain Monte Carlo (MCMC) \citep{Director17} to explore the low surface pressures for possible water and CO$_{2}$ atmospheres in rocky planets.

We describe the basics of our interior model, MSEI, in Sect. \ref{sec:interior_mod}. We explain the updates implemented in our atmosphere model, MSEIRADTRAN, with respect to similar previous k-correlated models \citep{Marcq17, Pluriel19} in Sect. \ref{sec:atm_mod}. In Sect. \ref{sec:mcmc}, we detail the implementation of the adaptive MCMC, and we show an example of the retrieval with it as well as with a non-adaptive MCMC.  With our model, we assess the observability of two planets that have been proposed for observations in Cycle 1 of JWST: TRAPPIST-1 c \citep{Gillon16,Grimm18} and 55 Cancri e \citep{Ehrenreich12,Bourrier18}. In Sect. \ref{sec:obs_data}, we summarise the planet and instrument parameters we use as input for our interior-atmospheric analyses and Pandexo \citep{Batalha20} to predict uncertainties in JWST observations. We present our results and conclusions in Sect. \ref{sec:result} and Sect. \ref{sec:conclusion}, respectively.



\section{Interior structure model}
\label{sec:interior_mod}


In this section, we review the fundamental principles on which our interior structure model is based. The input of the interior structure model are the total mass and two compositional parameters: the CMF and WMF. The CMF is defined as the mass of the Fe-rich core divided by the total planetary mass, while the WMF is the mass of the hysdrosphere divided by the total planetary mass. In the 1D interior model, the planetary radius, $r$, is represented by a 1D grid. Along this grid, the pressure, $P(r)$, the temperature, $T(r)$, the gravity acceleration, $g(r)$, and the density, $\rho(r)$, are calculated at each point. These four variables were obtained by solving the corresponding equation that defines its behaviour. The pressure was computed by integrating the equation of hydrostatic equilibrium (see Eq. \ref{eqn:dpdr}), while the temperature required integrating the adiabatic gradient profile (Eq. \ref{eqn:dtdr}). In low-mass planets, the opacity in their deep interior is high enough for the radiative temperature gradient to be greater than the adiabatic gradient, making the layers unstable against convection, according to the Schwarzschild criterion. In Eq. \ref{eqn:dtdr}, $\gamma$ and $\phi$ correspond to the Grüneisen and seismic parameters, respectively. The former describes the behaviour of the temperature in a crystal relative to its density. The latter parameter provides the speed at which seismic waves propagate in the same crystalline structure. Their formal definitions are shown in Eq. \ref{eqn:gruneisen}, where the seismic parameter can be seen as being related to the slope of the density at constant pressure, while the Grüneisen parameter depends on the derivative of the pressure with respect to the internal energy, $E$. The acceleration of gravity was obtained by solving the integral that results from Gauss's theorem (Eq. \ref{eqn:dgdr}), where $G$ is the gravitational constant and $m$ corresponds to the mass at a given radius, $r$.

\begin{equation}
\label{eqn:dpdr}
\dfrac{dP}{dr} = - \rho g,
\end{equation}

\begin{equation}
\label{eqn:dtdr}
\dfrac{dT}{dr} = - g \dfrac{\gamma T}{\phi},
\end{equation}

\begin{equation}
\label{eqn:gruneisen}
\begin{cases}
\phi = \dfrac{dP}{d \rho}  \\
\gamma = V \  \left(  \dfrac{dP}{dE} \right)_{V}, 
\end{cases}
\end{equation}

\begin{equation}
\label{eqn:dgdr}
\dfrac{dg}{dr} = 4 \pi G \rho - \dfrac{2 G m}{r^{3}}.
\end{equation}

The density, $\rho(r)$, was computed with the equation of state (EOS), which provides the density as a function of temperature and pressure. The interior structure model was divided into three separate layers: an Fe-rich core, a mantle rich in silicates, and a water layer. We used a different EOS to calculate the density for each of these layers. We adopted the Vinet EOS \citep{Vinet1989} with a thermal correction for the core and the mantle. More details about this EOS and its reference parameter values for the core and mantle can be found in \cite{Brugger16,Brugger17}. For the hydrosphere, we used the EOS and specific internal energy of \cite{Mazevet19} for supercritical and plasma phases of water, which is valid within the pressure and temperature regime ($P > 300$ bar, $T > 700$ K) covered by our interior structure model. We discuss the validity ranges of different water EOS for this regime in \cite{Acuna21}, while a detailed comparison of different EOS for high-pressure and high-temperature water and their effects on the total radius of the planet can be found in \cite{Aguichine21}. 

The final input for our interior structure model were the surface temperature and pressure. Together with the gravitational acceleration at the centre of the planet, whose value is zero, $g(r=0)$ = 0, these are the boundary conditions. Finally, the mass of each planetary layer was obtained by integrating the equation of conservation of mass (Eq. \ref{eqn:dmdr}). The total planetary mass is the sum of the individual mass of the layers. When the total input mass and the initial boundary conditions are met, the model reaches convergence. 

\begin{equation}
\label{eqn:dmdr}
\dfrac{dm}{dr} = 4 \pi r^{2} \rho.
\end{equation}

\subsection{Interior-atmosphere coupling}

The surface pressure for the interior model depends on the atmospheric mass on top of the outermost interface of the interior model. For envelopes whose bottom pressure is greater than or equal to $P$ = 300 bar, the interior model's surface pressure is set constant to 300 bar, which is the interface at which the interior and the atmosphere are coupled. Then the supercritical water layer extends from this interface to the boundary between the hydrosphere and the silicate mantle at higher pressures. For atmospheres whose surface pressure is less than 300 bar, the interior and atmosphere are coupled at the atmosphere-mantle interface, having the water envelope in vapour phase only. The WMF takes into account the mass of the atmosphere, $M_{atm}$. The atmospheric mass is calculated as shown in Eq. \ref{eqn:atm_mass}, where $P_{base}$ is the pressure at the base of the atmosphere (atmosphere-interior interface), $R_{bulk}$ is the radius from the centre of the planet to the base of the atmosphere, and $g_{surf}$ is the acceleration of gravity at this interface. The coupling interface between the interior and the atmosphere models at a maximum pressure of 300 bar is sufficiently close to the critical point ($P$ = 220 bar) of water to prevent the atmospheric model from taking over pressures at which convection dominates over radiation. The EOS we used for the interior \citep{Mazevet19} and the atmosphere \citep{AQUA} are based on the IAPWS-95 EOS. \cite{Wagner2002} report that the IAPWS-95 EOS presents unsatisfactory features in a small pressure and temperature range around the critical point. To prevent discontinuities in the adiabat and the density between the interior and the atmosphere for planets whose adiabat passes through this area, we set the coupling interface at 300 bar, not at $P_{crit}$ = 220 bar.

\begin{equation} \label{eqn:atm_mass}
    M_{atm} = \dfrac{P_{base} 4 \pi R_{bulk}^{2}}{g_{surf}}.
\end{equation}

The atmospheric model calculates the outgoing longwave radiation (OLR) and the Bond albedo, $A_{B}$, given as a function of bulk mass, radius, and temperature at the bottom of the atmosphere. If an atmosphere is in radiative equilibrium, its absorbed flux, $F_{abs}$ must be equal to its emitted radiation, which is the OLR. The absorbed flux depends on the Bond albedo via Eqs. \ref{eqn:rad} and \ref{eqn:teq}, where $\sigma$ is the Stefan-Boltzmann constant, and $T_{eq}$ is the planetary equilibrium temperature. This requires knowledge of the semi-major axis of the planet, $a_{d}$, as well as the stellar radius and effective temperature, $R_{\star}$ and $T_{\star}$, respectively.

\begin{equation} 
\label{eqn:rad}
F_{abs} = \sigma \ T_{\mathrm{eq}}^{4},
\end{equation}

\begin{equation} 
\label{eqn:teq}
T_{\mathrm{eq}} = (1-A_{B})^{0.25} \left( 0.5 \frac{R_{\star}}{a_{d}}\right)^{0.5} T_{\star}.
\end{equation}

For a constant planetary mass and radius, the temperature at the base of the atmosphere can be found by solving OLR$(T_{base})-F_{abs}(T_{base})=0$ with a root-finding method, such as the bisection method. Then, this root is the input boundary condition for the interior structure model. The radius calculated by the interior structure model (from the centre of the planet up to the base of the atmosphere) is an input for the atmospheric model, while the temperature at the bottom of the atmosphere is an input for both the interior and the atmospheric model. Therefore, the self-consistent coupling of both models is not straightforward and requires an iterative algorithm that checks that convergence is reached for the total radius and surface temperature. The total radius was computed as the sum of the bulk radius calculated by the interior model, and the atmospheric thickness was obtained by the atmospheric model. We refer the reader to \cite{Acuna21} for a detailed description of this algorithm.

\section{Atmospheric model}
\label{sec:atm_mod}


The interior-atmosphere coupling presented in our previous work \citep{Mousis20,Acuna21,Acuna22} was done by using grids of data generated by the atmospheric model of \cite{Pluriel19}. These grids provide the OLR, Bond albedo, and atmospheric thickness for a given set of mass, radius, and surface temperature when assuming a constant surface pressure. However, the use of these grids presents the following disadvantage:  The grids do not enable us to generate emission spectra that could be used to simulate observations. Therefore, we developed our own atmospheric model, MSEIRADTRAN. We started the development of MSEIRADTRAN by modifying the atmosphere model presented in \cite{Marcq17}\footnote{\url{http://marcq.page.latmos.ipsl.fr/radconv1d.html}} to include up-to-date opacity and EOS data. In the following, we summarise the basic structure and principles of MSEIRADTRAN and the atmospheric models presented in \cite{Marcq17,Pluriel19}.



We considered two scenarios for the composition of the envelope: water-dominated envelopes (99\% water plus 1\% CO$_{2}$) and CO$_{2}$-dominated envelopes (99\% CO$_{2}$ plus 1\% water). Including a wider variety of relative mass fractions between water and CO$_{2}$ in our models would only increase the degeneracies between atmospheric mass and atmospheric composition. Therefore, we only considered the two end-members to assess the observability of water and CO$_{2}$ spectral features with JWST. We did not model 100\% pure water or CO$_{2}$ atmospheres because such pure compositions are very unlikely due to outgassing and atmospheric escape \citep{KrissansenTotton21}. In addition, to make our comparison between MSEIRADTRAN and the atmospheric model of \cite{Pluriel19} consistent (see Fig. \ref{fig:MRdiag_MSEI}), we used the same exact compositions of 99\%:1\% instead of 100\% pure water or CO$_{2}$.

The 1D atmospheric model first proposes a pressure-temperature (PT) profile. This profile consists of a near-surface, dry convective layer followed by a wet convective region where condensation takes place and an isothermal mesosphere on top. If the surface temperature is cold enough to allow for condensation of water, the dry troposphere does not exist. For the isothermal mesosphere, we assumed a constant temperature of 200 K \citep{Marcq12,Marcq17}. The OLR is not very dependent on the temperature of an upper mesosphere \citep{Kasting88}. In addition, we did not take into account mesospheric stellar heating, which could significantly increase the temperature of the mesosphere. Therefore, adopting a low mesospheric temperature yields similar thermal profiles to self-consistent atmospheric calculations \citep{Lupu14}. The 1D grid that represents the pressure contains 512 computational layers. The adiabatic gradient used to calculate the temperature in each of these points in the convective regions depends on whether it is located in the dry or wet convective layer. The details of the computation of the wet and dry adiabatic gradients are presented in Sect. \ref{sec:atm_eos}.


The calculations of the emission spectrum and the Bond albedo were performed by bands. We divided the spectrum from 0 to 10100 cm$^{-1}$ (equivalent to $ \geq 1 \ \mu$m in wavelength) into 36 bands to obtain the OLR, similar to \cite{Pluriel19}. For each band, we calculated the total optical depth in each computational layer, which has four different contributions. These contributions are the optical depth due to collision-induced absorption (CIA) and line opacity (see Sect. \ref{sec:opacity}), Rayleigh scattering, and clouds. We treated Rayleigh scattering as was done in \cite{Pluriel19}, where the Rayleigh scattering opacity is related to wavelength, $\lambda$, following Eq. \ref{eq:Rayleigh}. The parameters $\kappa_{0}$ and $\lambda_{0}$ were adopted from \cite{Koppa13} and \cite{Sneep05} for H$_{2}$O, and CO$_{2}$, respectively. The opacity of clouds was considered for the atmospheric layers where condensation takes place. Similar to \cite{Marcq17} and \cite{Pluriel19}, the cloud opacity was parameterised after \cite{Kasting88}, who assumed a cloud opacity proportional to the extinction coefficient, $Q_{ext}$ (see Eq. \ref{eq:clouds}). The dependence of the extinction coefficient on wavelength (Eq. \ref{eq:qext}) is similar to that of water clouds on Earth \citep{Kasting88,Marcq17}.

\begin{equation} \label{eq:Rayleigh}
\kappa_{Rayleigh} (\lambda) = \kappa_{0}  \left(   \dfrac{\lambda_{0}}{\lambda}   \right)^{4},
\end{equation} 

\begin{equation} \label{eq:clouds}
\kappa_{clouds} (\lambda) = 130 \ Q_{ext} (\lambda), 
\end{equation}

\begin{equation} \label{eq:qext}
Q_{ext} = \begin{cases}
                        1 & \lambda \leq 20 \ \mu m \\
            3.26 \cdot \lambda^{-0.4} & \lambda > 20 \ \mu m. 
                 \end{cases}
\end{equation}

The total optical depth (Eq. \ref{eq:total_optd}) together with the PT profile are the input for the radiative transfer equation solver, DISORT \citep{DISORT}. DISORT obtains the emitted upward flux at the top of the atmosphere (TOA). The TOA flux was calculated for all 36 bands, which were then summed to obtain the bolometric, wavelength-integrated TOA flux, or OLR. Immediately after the OLR was computed, we started the calculation of the reflection spectrum and the Bond albedo, described in Sect. \ref{sec:albedo}.

\begin{equation} \label{eq:total_optd}
\tau_{total} = \tau_{line} + \tau_{Rayleigh} + \tau_{CIA} + \tau_{clouds}.
\end{equation}

\subsection{Atmospheric equation of state}
\label{sec:atm_eos}

The thermal structure of the atmosphere is divided into two main layers. The near-surface layer is adiabatic, which means convection takes place, while the layer on top is a mesosphere with a constant temperature. We set this temperature to 200 K, which is representative of the cool temperatures that hot low-mass planets present in their mesospheres \citep{Lupu14,Leconte13}. The adiabatic layer is divided into two sub-layers: a dry convective layer and a wet convective layer. Condensation may take place depending on the value of the pressure compared to the water saturation pressure. We considered the following envelope compositions: 1) 99\% water and 1\% CO$_{2}$ and 2) 1\% water and 99\% CO$_{2}$. Water is a condensable species, while CO$_{2}$ is a non-condensable gas. To obtain the temperature in a computational layer, $i$, we considered two approximations. The first is that the change in temperature and pressure within an atmospheric layer is small enough to approximate 
$\left( \dfrac{\partial T}{\partial P} \right)_{S} \simeq  \dfrac{T_{i-1}-T_{i}}{P_{i-1}-P_{i}}$. The second approximation is $\Delta P \sim P_{i} \ \Delta ln (P)$ because $\dfrac{d \ ln(P)}{dP} \sim \dfrac{\Delta ln(P)}{\Delta P} = \dfrac{1}{P_{i}}$. The temperature of a given computational layer, $T_{i}$, was then calculated as a function of the pressure in that layer, $P_{i}$, and the derivative of the temperature with respect to pressure at constant entropy, $( \partial T\partial P )_{S} $, as seen in Eq. \ref{eq:PT_final}.

\begin{equation} \label{eq:PT_final}
T_{i} = T_{i-1} - \left( \dfrac{\partial T}{\partial P} \right)_{S} \ P_{i} \ \Delta ln(P_{i}).
\end{equation}

If the pressure of water vapour is below the gas saturation pressure, $P_{v} < P_{sat}(T)$, or if its temperature is above the temperature of the critical point of water, $T > T_{crit}$, we are under dry convection \citep{Marcq12,Marcq17}. We calculated the derivative $(dT/dP)_{S}$ in the dry case following \cite{Marcq17} (Eq. \ref{eq:adiab_dry}). The densities of water vapour and CO$_{2}$ are $\rho_{v}$ and $\rho_{c}$, and $C_{p,v}$ and $C_{p,c}$ are their heat capacities. The specific volume of water vapour is indicated by $V_{v} = 1/ \rho_{v}$ .

\begin{equation} \label{eq:adiab_dry}
\left( \dfrac{\partial T}{ \partial P} \right) _{S, \ dry} = \dfrac{\rho_{v} \ T \ (\partial V_{v}/ \partial T)_{P} }{\rho_{v} \ C_{p,v} + \rho_{c} \ C_{p,c}}.
\end{equation}

When the atmospheric pressure reaches the water saturation curve, $P = P_{sat}(T)$, water vapour starts to condense and clouds form. Since the phase change requires energy in the form of latent heat, the wet adiabatic coefficient is different from the dry adiabatic one (Eq. \ref{eq:adiab_dry}). The expression for the derivative $(dT/dP)_{S}$ in the wet case is provided in Eq. \ref{eq:adiab_wet} \citep{Marcq17}. We note that CO$_{2}$ is the only non-condensable gas. The molecular weight of carbon dioxide is $M_{c}$, and $C_{v,c}$ is the specific heat capacity at a constant volume of CO$_{2}$. The ideal gas constant is $R$, while $\alpha_{v}$ is the mixing ratio of the water vapour density relative to CO$_{2}$, $\alpha_{v} = \rho_{v} / \rho_{c}$. Its derivative, $\partial \ ln (\alpha_{v}) / \partial \ ln (T)$ (Eq. \ref{eq:dalpha}), needed to be calculated as well \citep{Kasting88,Marcq12,Marcq17}.




\begin{multline} 
\left( \dfrac{\partial T}{ \partial P} \right) _{S, \ wet} = \\
\dfrac{1}{ (dP_{sat}/dT) + \rho_{c}R/M_{c} (1+ \partial \ ln (\rho_{v}) / \partial \ ln (T) - \partial \ ln (\alpha_{v}) / \partial \ ln (T))  }
\end{multline}



\begin{equation} \label{eq:dalpha}
\dfrac{\partial \ ln (\alpha_{v})}{\partial \ ln (T)} = \dfrac{R/M_{c}(\partial \ ln (\rho_{v}) / \partial \ ln(T)) - C_{v,c} - \alpha_{v} (\partial s_{v} / \partial \ ln(T))}{\alpha_{v} (s_{v}-s_{c}) + R/M_{c}}.
\end{equation}

\begin{figure}[h]
   \centering
   \includegraphics[width=\hsize]{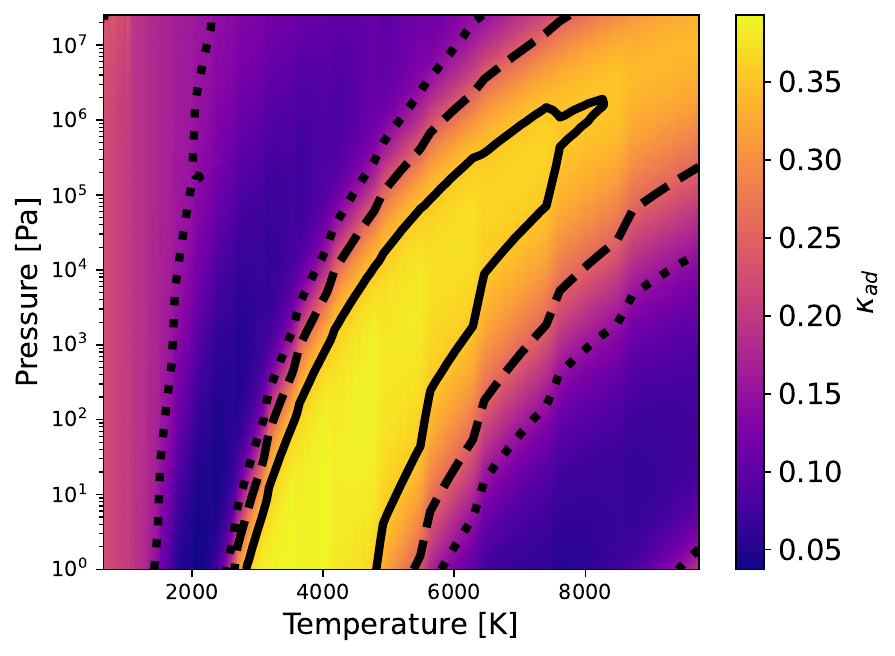}
      \caption{Dry adiabatic coefficient for water, $\kappa_{ad}$, as a function of pressure and temperature. The displayed region covers the cold and hot gas phase of water up to pressures close to the critical point. Solid, dashed, and dotted contours correspond to $\kappa_{ad}$ = 0.35, 0.25, and 0.15, respectively.}
         \label{fig:adiab_dry}
\end{figure}

The density and heat capacity of water were previously obtained by using the steam tables of \cite{Haar84}. These tables treat water as a non-ideal gas, although they are not valid for $T > 2500$ K. Therefore, for temperatures higher than 2500 K, we used the EOS tables from \cite{AQUA} to calculate the thermodynamic properties of water. These tables are a compilation of different EOSs, where each EOS is applied in its validity region of the water phase diagram. There are two EOSs that are used in the region relevant for the atmospheres of low-mass, highly irradiated planets. The first EOS is the IAPWS95 \citep{Wagner2002}, whose validity range for the high-pressure supercritical regime corresponds to 251 to 1273 K in temperature and up to 1 GPa in pressure. \cite{AQUA} transitioned at 1200 K to an EOS that is valid at low pressures and high temperatures. This second EOS is the CEA (Chemical Equilibrium with Applications) package \citep{Gordon_cea,Mcbride_cea}. This package incorporates the effects of single ionisation and thermal dissociation, which are processes that occur only at high temperatures in the gas phase. Figure \ref{fig:adiab_dry} shows the dry adiabatic coefficient as a function of pressure and temperature in the region of the water phase diagram relevant for hot planetary atmospheres. The temperature derivative $(dT/dP)_{S}$ is closely related to the adiabatic coefficient, $\kappa_{ad}$ (Eq. \ref{eq:kappadef}). The reduction of the dry adiabatic coefficient at $T = $ 1000 to 2500 K is due to thermal dissociation, whereas the decrease at higher temperatures ($T \geq$ 6000 K) is caused by thermal ionisation \citep{AQUA}.

\begin{equation} \label{eq:kappadef}
\left(  \dfrac{\partial T}{\partial P} \right) _{S} = \dfrac{T}{P} \ \kappa_{ad} (P,T).
\end{equation}

We assumed that CO$_{2}$ is an ideal gas to calculate its density, $\rho_{c}$. As we treat CO$_{2}$ as an ideal gas, we can calculate its heat capacity $C_{v,c} = C_{p,c} - R/M_{c}$. We calculated the specific heat of CO$_{2}$ by using Shomate's equation (Eq. \ref{eq:cpc}), whose tabulated coefficients $A$ to $E$ are provided by \cite{Chase98}. This is part of the National Institute of Standards and Technology (NIST) chemistry WebBook database\footnote{\url{https://webbook.nist.gov/cgi/cbook.cgi?ID=C124389&Mask=1}}, and it is valid for temperatures up to 6000 K.

\begin{equation} \label{eq:cpc}
C_{p,c}(T) = A + B \ T + C \ T^{2} + D \ T^{3} + E/T^{2}.
\end{equation}

Finally, we determined the atmospheric thickness, $z_{atm}$, under the assumption of hydrostatic equilibrium (Eq. \ref{eqn:dpdr}). In the case of the atmospheric model, the pressure is integrated over altitude, $z$, instead of internal radius, $r$. The altitude of the computational atmospheric layer $i$ is shown in Eq. \ref{eq:altitude}, where $\rho_{total, i}$ is the total mass density at altitude $z_{i}$, $\rho_{total,i} = \rho_{v, i} + \rho_{c, i}$. This expression was derived by approximating $(d P/d z) \sim \Delta P/z_{i-1}-z_{i}$. The gravity acceleration was computed at each point of the 1D grid with the equation for hydrostatic equilibrium. This is noted explicitly in Eq. \ref{eq:altitude} by the labelling of the gravity at the computational layer $i$ as $g_{i}$. The final atmospheric thickness was obtained by evaluating $z$ at the transit pressure, 20 mbar. 

\begin{equation} \label{eq:altitude}
z_{i} = z_{i-1} + \dfrac{P_{i} \ \Delta ln(P)}{g_{i} \ \rho_{total, i}}.
\end{equation}

\subsection{K-correlated method implementation}
\label{sec:k-corr_met}

We employed the k-correlated method \citep{Liou80} to compute the emission spectrum. We discretised the dependence of the opacity on wavenumber, $\nu$, by dividing the spectrum into bands. The spectral transmittance of a spectral band $b$, $\mathcal{T}_{b}$, is defined as the exponential sum of the opacity of the band, $\kappa^{b}$, times the column density $m$, which is only dependent on pressure and temperature \citep{sanchezlavega}. The exponential sum was performed over $G$-points, which are the abscissa values chosen to discretise the cumulative probability function of the opacity, $G(\kappa)$. Each $G$-point, $G_{i}$, has an associated weight in the exponential sum in Eq. \ref{eq:optimized}, $w_{i}$. The discrete opacity value is not only dependent on the spectral band but also on the $G$-point, which is noted explicitly in Eq. \ref{eq:optimized} as $\kappa^{b}_{i}$. The total number of $G$-points is indicated by $N_{G}$.

\begin{equation} \label{eq:optimized}
\mathcal{T}^{b} (m) = 
 \int_{0}^{1} e^{-\kappa^{\nu}(G) \ m} \ dG \simeq \sum_{i=1}^{N_{G}} e^{-\kappa^{b}_{i} \ m(P,T)} \ w_{i}. 
\end{equation}

\begin{equation} \label{eq:flux_band}
F^{\uparrow}_{TOA,\ b} =  \int_{0}^{1} F^{b}(G) \ dG = \sum_{i=1}^{N_{G}} F^{b}_{G_{i}} \ w_{i}.
\end{equation}

In the case of an arbitrary set of $G$-points, the weights are equal to $w_{i} = \Delta G_{i}$, which are the widths of the bins in the $G$-space. In each atmospheric layer, the pressure and temperature are considered constant. Under this condition and within each spectral band, we can exchange wavenumber with $G$ \citep{Mollierephd}. Consequently, we integrated the upward flux over $G$ to obtain the emitted flux within each band (Eq. \ref{eq:flux_band}). The upward top flux per bin and per $G$-point, $F^{b}_{G_{i}}$, was obtained by invoking the radiative solver for a total optical depth whose line optical depth is $\tau_{line} = \tau^{b}_{i}$ (Eq. \ref{eq:tau_gpoint}).

\begin{equation} \label{eq:tau_gpoint}
\tau^{b}_{i} = \kappa^{b}_{i} \ \rho \ \Delta z.
\end{equation} 

Following \cite{Malik17,Malik19}, we defined our discrete $G$-points as the roots of the Legendre polynomial, $G_{LG,i}$ (Eq. \ref{eq:lg_gpoints}). The corresponding weigths, $w_{i}$, are the Legendre-Gaussian (LG) weights associated to the Legendre polynomial of the $N_{G}$th order, $P_{N_{G}}$ (Eq. \ref{eq:weights}). The LG weights were calculated from the $i$th root of the $N_{G}$th order Legendre polynomial, $y_{i}$, as well as from the polynomial's derivative, $P'_{N_{G}}$ \citep{Abramowitz72,Malik17}. We chose to have $N_{G} =$ 16 $G$-points.


\begin{equation} \label{eq:lg_gpoints}
G_{LG,i} = \dfrac{ ( 1 + y_{i} ) }{2},
\end{equation}


\begin{equation} \label{eq:weights}
w_{LG,i} = \dfrac{2}{\left( 1 - y_{i}^{2} \right) P'_{N_{G}} \left(  y_{i} \right)^{2} }.
\end{equation}


To generate emission spectra and assess the observability with JWST, we used the original spectral resolution of the opacity k-table data (see Sect. \ref{sec:opacity}) in our k-correlated model. This spectral resolution corresponds to $R = 200$ to 300 in the spectral range $\lambda$ = 1 to 20 $\mu$m.

\subsection{Opacity data}
\label{sec:opacity}

The total optical depth computed in the atmospheric model includes contributions from collision-induced absorption (CIA) and line absorption. The CIA absorption is particularly important in dense gases, such as steam and CO$_{2}$ at high pressures, especially if the line opacity is weak \citep{Pluriel19}. For the atmospheric compositions we considered in this work, namely 99\% H$_{2}$O:1\% CO$_{2}$ and 99\% CO$_{2}$:1\% H$_{2}$O, we required CIA data for collisions between these two gases and their self-induced absorptions. We adopted CIA absorption data for H$_{2}$O-CO$_{2}$ and H$_{2}$O-H$_{2}$O collisions provided by \cite{Ma92} and \cite{TurbetCIA}\footnote{\url{https://www.lmd.jussieu.fr/~lmdz/planets/LMDZ.GENERIC/datagcm/continuum_data/}}, respectively. The CO$_{2}$-CO$_{2}$ CIA opacities are from a look-up table obtained by \cite{Bezard11} and \cite{Marcq08}, which is also used in the atmospheric model by \cite{Marcq17}. Our H$_{2}$O-H$_{2}$O CIA table covers the complete spectral range where we calculated both our emission and reflection spectra, while the H$_{2}$O-CO$_{2}$ CIA table covers the bands with wavelength $\geq$ 1 $\mu$m, which corresponds to the bands necessary for the emission spectrum only. For the bands whose wavelengths are outside the spectral range of the CIA table, we assumed a constant CIA opacity value equal to the opacity at the limit band of the table.

\cite{heliosk} provide a database\footnote{\url{https://chaldene.unibe.ch/data/Opacity3/}} of pre-calculated opacity k-tables for different species and line lists. For water- and CO$_{2}$-dominated atmospheres, we adopted the POKAZATEL \citep{pokazatel} and HITEMP2010 \citep{HITEMP} opacity data, respectively. POKAZATEL presents the widest validity range in temperature for water in planetary atmospheres, with a maximum temperature of 5000 K, while the HITEMP maximum temperature is 4000 K. Following the procedure described in \cite{Leconte21}, we binned the k-correlated opacities to the same spectral bins of \cite{Marcq17} and \cite{Pluriel19}. We calculated the k-coefficients for our water-CO$_{2}$ mixture by assuming that the spectral features of the individual gases are correlated \citep{Malik17}. In the correlated approximation, the mixed opacity was estimated as indicated in Eq. \ref{eq:corr_mix}, where $\chi_{j}$ is the mixing ratio by mass of the $j$th gas and  $\kappa_{j, i}$ is the k-coefficient of the $j$th gas evaluated at the $G_{i}$ point. The mixing ratio by mass is defined as $\chi_{j} = \dfrac{X_{j} \ MW_{j}}{\mu}$, where $X_{j} = \dfrac{P_{j}}{P}$ is the volume mixing ratio of the $j$th species, $MW_{j}$ is its molecular weight, and $\mu$ is the mean molecular weight of the mixture.

\begin{equation} \label{eq:corr_mix}
\kappa_{mix, i} = \sum_{j=1}^{N_{gases}} \chi_{j} \ \kappa_{j, i}.
\end{equation}

\subsection{Reflection spectra and Bond albedo}
\label{sec:albedo}

Once the bolometric OLR was obtained, we initiated the calculation of the reflectivity in 30 bands, from 5 to 0.29 $\mu$m, to obtain the Bond albedo \citep{Pluriel19}. The bands for which we calculated both the emission flux and the reflectivity (from 1 to 5 $\mu$m) could not have the two quantities calculated simultaneously since DISORT requires different input settings to calculate them. For the emission, we assumed zero illumination from the top of the atmosphere, as well as an upward flux that forms 90 degrees with the surface of the planet,  which corresponds to a polar angle equal to zero. DISORT calculates the reflectivity of the atmosphere as a function of incident beam angle, which corresponds to the solar zenith angle (SZA), $\theta$ in Eq. \ref{eq:bond}. This is the angle that the incident light forms with the normal of the incident surface. Once we obtained the dependence of the reflectivity with SZA, we could average it as indicated in Eq. \ref{eq:bond} \citep{Simonelli88}. To integrate Eq. \ref{eq:bond}, we evaluated the reflectivity at ten different SZA values. We assumed four streams for DISORT in both the calculation of the OLR and the reflectivity.
  
\begin{equation} \label{eq:bond}
A_{B}(\nu) = 2  \int_{0}^{\pi/2} A_{B}(\nu,\theta) \ cos (\theta) \ sin(\theta) \ d\theta.
\end{equation} 

After averaging the reflectivity over SZA, we obtained the reflection spectrum, which is the dependence of the albedo as a function of the wavenumber. To obtain the bolometric Bond albedo, we integrated Eq. \ref{eq:alb_bol} \citep{Pluriel19}. The variable $A_{B}(\nu)$ is the reflectivity as a function of the wavenumber, $B_{\nu}(T_{\star})$ is Planck's function for a temperature equal to the effective temperature of the host star, and $\sigma$ is the Stefan-Boltzmann constant.

\begin{equation} \label{eq:alb_bol}
A_{Bond, \ bol} = \dfrac{\pi \ \int_{0}^{\infty}  A_{B}(\nu) \ B_{\nu}(T_{\star})  \ d \nu  }{\sigma T_{\star}^{4}}.
\end{equation} 

The Bond albedo is a parameter particularly sensitive to the choice of phase function. For atmospheric layers that present clouds, the gas contributes to scattering with a Rayleigh phase function, while we assumed the Henyey-Greenstein phase funtion for clouds. DISORT requires the calculation of the Legendre moments of the combined phase function, which we estimated as the weighted average of the moments of the two individual phase functions \citep{Liou80,Boucher98}. The weights were calculated as the ratio of the optical depth due to Rayleigh scattering or clouds divided by the total optical depth, $\tau_{Rayleigh} + \tau_{clouds}$, for Rayleigh and Henyey-Greenstein phase function moments, respectively. For clear atmospheric layers, the only contribution to scattering is Rayleigh scattering due to the gas, so the total phase function moment corresponds to that of Rayleigh scattering \citep{Marcq17}.

As input, DISORT also requires the single scattering albedo of each atmospheric layer. The single scattering albedo is defined as the ratio of scattering efficiency to total extinction efficiency. The total extinction is a sum of both extinction by scattering and extinction by absorption. Therefore, a single scattering albedo of one indicates that all extinction is due to scattering, whereas a value of zero means that absorption dominates. Similar to the moments of the phase function, we estimated the combined single scattering albedo from gas (Rayleigh) and clouds with their weighted average. The single scattering albedo due to Rayleigh scattering was calculated as the Rayleigh optical depth divided by the total optical depth, $\tau_{clear} + \tau_{Rayleigh}$. The clear optical depth is the sum of the line and CIA optical depths (Sect. \ref{sec:opacity}). For fast computations of the Bond albedo within our interior-atmosphere model in retrievals, we use a grey model for the line opacity. The line opacity is constant with wavelength, being 0.01 m2/kg for H$_{2}$O. This grey opacity is benchmarked with non-grey atmospheric models \citep{Nakajima92,Marcq17}. In the case of water, a grey opacity of 0.01 m$^{2}$/kg is representative of the opacity of water in the 8-20 $\mu$m spectral window at the Standard Reference Point \citep{Ingersoll69}. The single scattering albedo due to clouds was calculated as the ratio of the clouds optical depth divided by $\tau_{clear} + \tau_{cloud}$, times the cloud single scattering albedo defined in \cite{Kasting88}:

\begin{equation}
   \varpi_{0} = 
    \begin{cases}
        1 & \lambda \leq 2 \ \mu m \\
        1.24 \cdot \lambda^{-0.32} & \lambda > 2 \ \mu m \\
    \end{cases}.
\end{equation}

\subsection{Comparison to previous models}
\label{sec:validation}

\begin{figure*}[h]
        \centering
        \includegraphics[width=\textwidth]{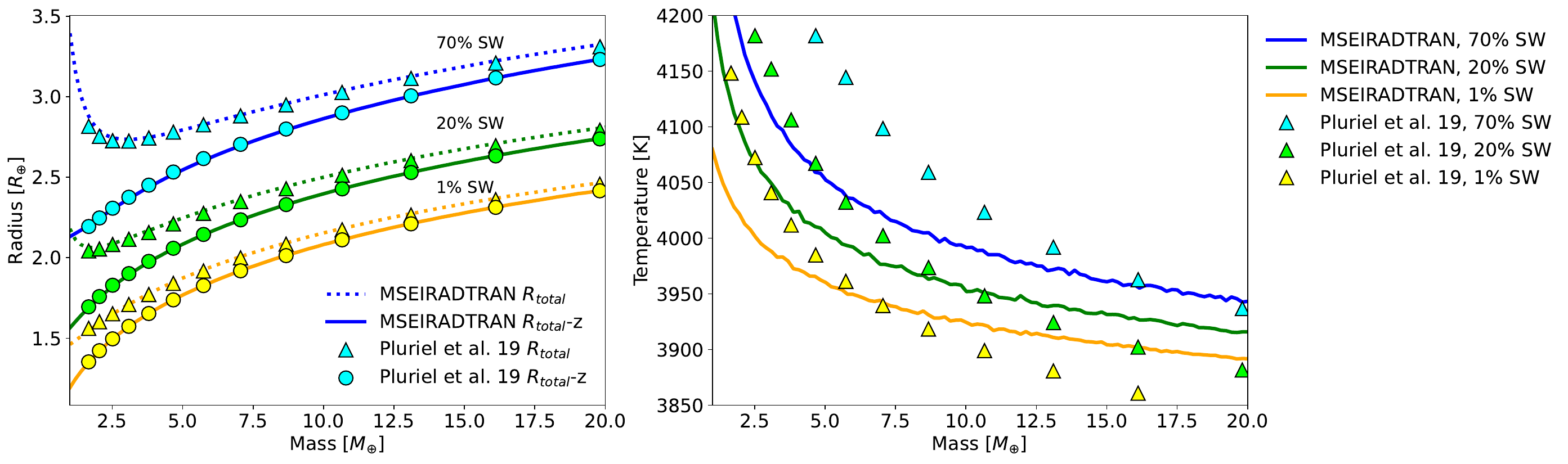}
        \caption{Comparison of the radius and interior-atmosphere boundary temperature between our k-correlated model (MSEIRADTRAN) and that of \cite{Pluriel19}. Left panel: Mass-radius relationships for a planet with a water-dominated atmosphere orbiting a Sun-like star at $a_{d} = 0.05$ AU. Dashed lines indicate the total radius calculated by MSEIRADTRAN, while the solid line corresponds to the interior radius, which comprises the core, mantle, and supercritical water (SW) hydrosphere. Triangles and circles indicate the total radius and the interior radius obtained when the interior model is coupled with the atmospheric model of \cite{Pluriel19}, respectively. Right panel: Temperature at the 300 bar interface as a function of planetary mass.}
        \label{fig:MRdiag_MSEI}
\end{figure*}

To compare the effect of the temperature at the interior-atmosphere boundary on the total radius of planets with water-rich envelopes, we computed two sets of mass-radius relationships (Fig. \ref{fig:MRdiag_MSEI}, left panel). The first set was obtained by coupling the interior model with our k-correlated model, and the second one was obtained with that of \cite{Pluriel19}. The difference in radius between the two models is less than 1\% in all masses and water contents. 
The difference in temperature between the two models is within 50 K for water mass fractions of 1\% and 20\% (Fig. \ref{fig:MRdiag_MSEI}, right panel). For WMF = 70\% and masses below 8 $M_{\oplus}$, the difference in temperature between the two models can reach up to 130 K. This discrepancy is caused by differences in the OLR between k-correlated models that use different opacity data. Nonetheless, planets with $M < 8 \ M_{\oplus}$ are unlikely to accrete water mass fractions above 50\% \citep{Miguel20,Kimura22}. This means that differences in opacity data in atmospheric models are unlikely to affect mass-radius relations and interior structure retrievals of detected exoplanets. We tested approximations to atmospheric models, including grey models, and found that differences in interface temperature greater than 150 K can produce changes in radius of more than 1\% (not shown).

\section{Markov chain Monte Carlo}
\label{sec:mcmc}


The Markov chain Monte Carlo (MCMC) Bayesian algorithm described in \cite{Dorn15} was later adapted by \cite{Acuna21} to our forward interior-atmosphere model. In this work, we use it to perform the retrievals. We recall that the model parameters are the planetary mass, $M$; the CMF, $x_{core}$; and the WMF, $x_{H2O}$. Therefore, one single model is determined by these three parameters as \textbf{m} = $\left\lbrace M, x_{core}, x_{H2O} \right\rbrace$. The available data are the total mass $M$, the total radius $R$, and the Fe/Si abundance, \textbf{d} = $\left\lbrace M_{obs}, R_{obs}, Fe/Si_{obs} \right\rbrace$, whose observational errors are $\sigma(M_{obs}),\sigma(R_{obs}),\sigma(Fe/Si_{obs})$, respectively. When the Fe/Si mole ratio is not considered in the inverse problem, the data is reduced to only the total planetary mass and radius, \textbf{d} = $\left\lbrace M_{obs}, R_{obs} \right\rbrace$. The prior information consists of a Gaussian distribution centred on the mean value of the observed mass with a standard deviation equal to the observational uncertainty. For the CMF and WMF, we considered uniform distributions ranging from zero to one as priors. The MCMC scheme starts by first drawing a value for each of the model parameters from their prior distributions, which we denote as \textbf{m}$_{1} = \left\lbrace M_{1},x_{core,1}, x_{H2O,1} \right\rbrace $. The interior model calculates the planetary radius and Fe/Si abundance that correspond to these model parameters, which is g(\textbf{m}$_{1}) = \left\lbrace R_{1},M_{1},Fe/Si_{1} \right\rbrace $. We then computed the likelihood of this model (Eqs. \ref{eq:likelihood} and \ref{eq:norm}), and we drew a new model from the prior distributions, $\textbf{m}_{2}$.

\begin{multline} \label{eq:likelihood}
L(\textbf{m}_{i} \mid \textbf{d}) = 
C \ exp \ \Biggl( -\frac{1}{2} \Biggl[
\left( \frac{(R_{i}-R_{obs})}{\sigma(R_{obs})}\right)^{2} 
+ \left( \frac{(M_{i}-M_{obs})}{\sigma(M_{obs})} \right)^{2}\\
+ \left( \frac{(Fe/Si_{i}-Fe/Si_{obs})}{\sigma(Fe/Si_{obs})} \right)^{2}
 \Biggr]  \Biggr), 
\end{multline}

\begin{equation} \label{eq:norm}
C = \dfrac{1}{(2 \pi)^{3/2} \left[ \sigma^{2}(M_{obs}) \cdot \sigma^{2}(R_{obs})  \cdot \sigma^{2}(Fe/Si_{obs}) \right]^{1/2} }.
\end{equation}

The log-likelihoods, $l(\textbf{m}_{i} \mid \textbf{d}) = log(L(\textbf{m}_{i} \mid \textbf{d}))$, of both models were used to estimate the acceptance probability (Eq. \ref{eq:acc_prob}). Consecutively, a random number was drawn from a uniform distribution between zero and one. If $P_{accept}$ was greater than this random number, \textbf{m}$_{2}$ was then accepted, and the chain moved to this set of model parameters,  starting the following chain $n+1$. Otherwise, the chain remained in \textbf{m}$_{1}$, and a different set of model parameters was proposed, \textbf{m}$_{3}$. The accepted models were stored, and the values of their parameters composed the PDF that would enable us to estimate their mean and uncertainties.

\begin{equation}
\label{eq:acc_prob}
P_{accept} = min \left\lbrace 1, e^{(l(\textbf{m}_{new,i} \mid \textbf{d}) - l(\textbf{m}_{old} \mid \textbf{d}))}  \right\rbrace .
\end{equation}

\subsection{Adaptive Markov chain Monte Carlo}

In our initial implementation of the MCMC \citep{Acuna21}, the random walker used a uniform distribution  to choose the next state where it would move in the parameter space of the CMF and WMF. This approach  is called a naive walk \citep{MT95}, and in it all the points in the parameter space have a probability of being chosen that is proportional to their number of neighbours. This poses the following problem: For the states whose CMF or WMF is close to zero or one, they are less likely to be sampled in the random walk because they have less neighbours than the central values. A model with a WMF equal to one is not physical, although many highly irradiated rocky planets might present low-mass atmospheres that correspond to a WMF close to zero. To compensate for this lower probability of being chosen in the limiting states of the prior, we used an adaptive step size in the walker. This consisted of having an adaptive maximum size for the perturbation used to generate a new model instead of using a fixed value everywhere in the parameter space. This adaptive step size would decrease in the limiting areas of the prior (i.e. low WMF states) and have its greatest value at the centre of the prior (WMF = 0.5). The self-adjusting step size can be carried by a transformation of the parameter space, which ranges from exponential to spherical transforms \citep{Chaudhry21}. In this work, we chose to implement the self-adjusting logit transform (SALT) proposed by \cite{Director17}. The SALT transform is publicly available in the \textit{SALTSampler} R package\footnote{\url{https://rdrr.io/cran/SALTSampler/man/SALTSampler-package.html}}, which eases its implementation in Python for our own model.

We compared the non-adaptive and adaptive MCMC for one planet, TOI-220 b \citep{Hoyer21}. We considered as input data, the total mass and radius as well as the Fe/Si mole ratio, which was calculated with the stellar abundances of the host star. We did not establish maximum limits for the CMF and the WMF. The planet TOI-220 b has an equilibrium temperature of 806 K, which means that it is strongly irradiated and could present steam and supercritical phases. Table \ref{tab:TOI220b} presents the input data and the retrieved parameters of the non-adaptive and adaptive MCMCs. All three agree within uncertainties for mass, radius, and Fe/Si. The uncertainties of the mass and radius in the non-adaptive MCMC are smaller than the input data. This difference in uncertainties is significant in the case of the total mass. This discrepancy in uncertainty indicates that the non-adaptive MCMC is not as effective as the adaptive MCMC at sampling all possible $\left\lbrace x_{core}, x_{H2O} \right\rbrace$ pairs that could reproduce the mass and radius data. As a consequence, the uncertainties of the WMF are underestimated in the non-adaptive MCMC, while the adaptive MCMC produces a greater confidence interval for the WMF and retrieves the exact uncertainties of the mass and radius.

\begin{table}[H]
\caption{TOI-220 b MCMC input (Data) and output mean values and 1$\sigma$ uncertainties for the non-adaptive and adaptive MCMCs.}
\centering
\begin{tabular}{cccc}
\hline \hline
                           & Data \citep{Hoyer21}          & Non-adaptive  & Adaptive      \\ \hline
\multicolumn{1}{c|}{$M$ [$M_{\oplus}$]}     & 13.8$\pm$1.0  & 13.8$\pm$0.7  & 13.7$\pm$1.0  \\
\multicolumn{1}{c|}{$R$ [$R_{\oplus}$]}     & 3.03$\pm$0.15 & 3.06$\pm$0.12 & 2.98$\pm$0.15 \\
\multicolumn{1}{c|}{Fe/Si} & 0.65$\pm$0.09 & 0.64$\pm$0.11 & 0.64$\pm$0.10 \\
\multicolumn{1}{c|}{$x_{core}$}   &               & 0.08$\pm$0.03 & 0.09$\pm$0.03 \\
\multicolumn{1}{c|}{$x_{H2O}$}   &               & 0.62$\pm$0.10 & 0.58$\pm$0.14 \\ \hline
\end{tabular}
\label{tab:TOI220b}
\end{table}

\begin{figure}[h]
  \centering
   \includegraphics[width=\hsize]{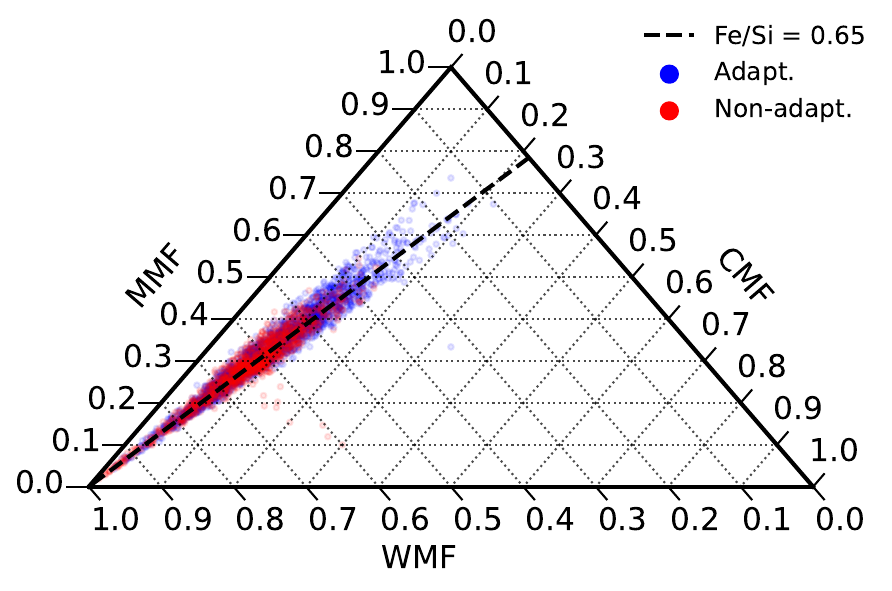}
      \caption{Sampled 2D PDFs of the CMF and WMF in the ternary diagram for TOI-220 b for non-adaptive (red) and adaptive (blue) MCMCs. The mean value of the input Fe/Si mole ratio is indicated with a dashed black line. The mantle mass fraction (MMF) is defined as MMF = 1 - CMF - WMF.}
         \label{fig:MCMC_comp}
\end{figure}

In Fig. \ref{fig:MCMC_comp}, we show the sampled 2D PDFs for the CMF and the WMF in the ternary diagram. In addition to the same area of the ternary diagram as the non-adaptive algorithm, the adaptive MCMC explores an area at lower WMF along the Fe/Si = 0.65 isoline, going down to WMF = 0.10 in the driest simulations. This is a consequence of the ability of the adaptive MCMC to better sample the extremes of the prior distribution of the WMF, in comparison to the non-adaptive MCMC. Furthermore, the acceptance rate is also improved in the adaptive case, having an acceptance rate of 53\% in comparison to the original acceptance rate of 35\% for the non-adaptive case within the same time.

\section{Planetary and observation parameters}
\label{sec:obs_data}


Our MCMC analysis required as input the planetary mass, radius, and Fe/Si mole ratio for the two planets we considered as test cases in this work, TRAPPIST-1 c and 55 Cancri e. Their values and references are shown in Table \ref{tab:planet_data}. To determine the surface temperature at which the atmosphere is in radiative-convective equilibrium, the stellar effective temperature, stellar radius, and semi-major axis were needed, as seen in Eq. \ref{eqn:teq}. The values we adopted and their references are given in Table \ref{tab:planet_data}.

\begin{table*}[h]
\caption{Planetary parameters for TRAPPIST-1 c and 55 Cancri e: masses, radii, Fe/Si mole ratios, semi-major axes, and host stellar effective temperatures and radii.}
\label{tab:planet_data}
\centering
\begin{tabular}{cccccccc}
\hline \hline
             & $M$ [$M_{\oplus}$] & $R$ [$R_{\oplus}$]& Fe/Si & $a_{d}$ [AU] & $T_{\star}$ [K] & $R_{\star}$ [$R_{\odot}$] & References \\ \hline
\multirow{2}{*}{TRAPPIST-1 c} & \multirow{2}{*}{1.308$\pm$0.056} & \multirow{2}{*}{1.097$^{+0.014}_{-0.012}$} & \multirow{2}{*}{0.76$\pm$0.12} & \multirow{2}{*}{1.58$\times \ 10^{-2}$} & \multirow{2}{*}{2566} & \multirow{2}{*}{0.119} & \multirow{2}{*}{1, 2}\\
 &  &  &  &  &  &  &  \\

\multirow{2}{*}{55 Cancri e} & \multirow{2}{*}{7.99$^{+0.32}_{-0.33}$} & \multirow{2}{*}{1.875$\pm$0.029} & \multirow{2}{*}{0.60$\pm$0.14} & \multirow{2}{*}{1.54$\times \ 10^{-2}$} & \multirow{2}{*}{5172} & \multirow{2}{*}{0.943} & \multirow{2}{*}{3, 4} \\
 &  &  &  &  &  &  &  \\ \hline 
\end{tabular}%
\tablebib{(1)~\citet{Agol21};
(2) \citet{Unterborn18};
(3) \citet{Bourrier18};
(4) \citet{Luck16}
.}
\end{table*}

To simulate observations with JWST in photometry, we assumed the atmospheric parameters retrieved in our adaptive MCMC analysis and generated emission spectra with their respective temperature-pressure profiles. Consecutively, we binned the emission spectrum using the response functions of each of the MIRI photometry filters\footnote{\url{http://svo2.cab.inta-csic.es/svo/theory/fps3/index.php?id=JWST}} \citep{Glasse15,Piette22}. The mean flux, $\langle f_{\lambda} \rangle$, of an emission spectrum, $ f(\lambda) $, observed with a filter with transmission function $R(\lambda)$, is defined in Eq. \ref{eq:filter_mean} \citep{Stolker20}. We considered a random Gaussian noise of 100 ppm for each filter in order to derive the uncertainties of the mean flux \citep{lustigyaeger19,Piette22}.

\begin{equation} \label{eq:filter_mean}
\langle f_{\lambda} \rangle = \dfrac{ \int f(\lambda) \ R(\lambda) \ d \lambda}{\int R(\lambda) \ d \lambda}.
\end{equation}


For the observation of the emission spectrum of 55 Cancri e, we used Pandexo \citep{Batalha20} to simulate the expected noise. Our input included the stellar effective temperature as well as the stellar and planet radius (see Table \ref{tab:planet_data}). Additional input parameters can be found in the database accessible by Pandexo and ExoMast, which are shown in Table \ref{tab:obs_param}. We adopted observation and instrumentation variables from \cite{Hu_jwstprop_cnc}. 

\begin{table}[h]
\caption{Input parameters for Pandexo to simulate observations of the emission spectrum of 55 Cancri e with JWST's MIRI LRS and NIRCam instruments.}
\resizebox{0.49\textwidth}{!}{%
\begin{tabular}{llcc}
\hline \hline
\multicolumn{2}{l}{Parameter} & \multicolumn{2}{c}{Value} \\ \hline
\multicolumn{2}{l}{\textit{Star}} &  \\
\multicolumn{2}{l}{Metallicity, $log$[Fe/H]} & \multicolumn{2}{c}{0.35} \\
\multicolumn{2}{l}{Gravity, $log \ g$ [cgs]} & \multicolumn{2}{c}{4.43} \\ 
\multicolumn{2}{l}{J Magnitude} & \multicolumn{2}{c}{4.59} \\ \hline
\multicolumn{2}{l}{\textit{Planet}} &  \\
\multicolumn{2}{l}{Transit duration [d]} & \multicolumn{2}{c}{0.0647} \\ \hline
\multicolumn{2}{l}{\textit{Observation}} &  \\
\multicolumn{2}{l}{Baseline [h]} & \multicolumn{2}{c}{3.2}  \\
\multicolumn{2}{l}{Number of eclipses} & \multicolumn{2}{c}{2} \\
\multicolumn{2}{l}{Instrument} & MIRI LRS & NIRCam \\ 
\multicolumn{2}{l}{Mode} & Slitless & F444W, subgrism 64 \\
\multicolumn{2}{l}{Saturation limit [full well]} & \multicolumn{2}{c}{80\%} \\
\multicolumn{2}{l}{Constant minimum noise} & \multicolumn{2}{c}{100 ppm} \\ \hline
\end{tabular}%
}
\label{tab:obs_param}
\end{table}


\section{Interior composition and simulated spectra}
\label{sec:result}


\subsection{TRAPPIST-1 c}

\begin{figure*}[h]
  \begin{center}
    \subfloat[H$_{2}$O]{
      \includegraphics[width=0.5\textwidth]{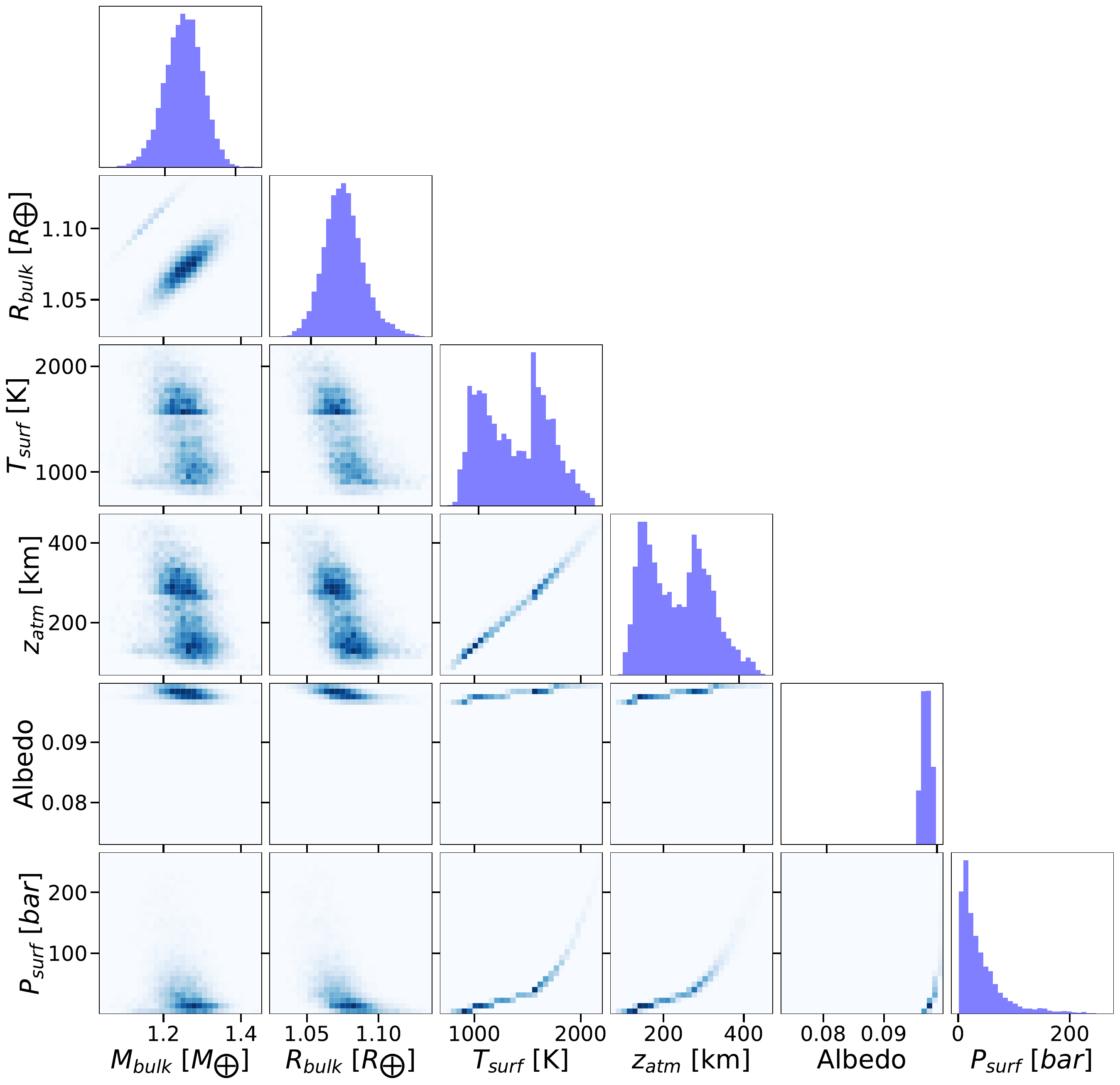}
                         }
    \subfloat[CO$_{2}$]{
      \includegraphics[width=0.5\textwidth]{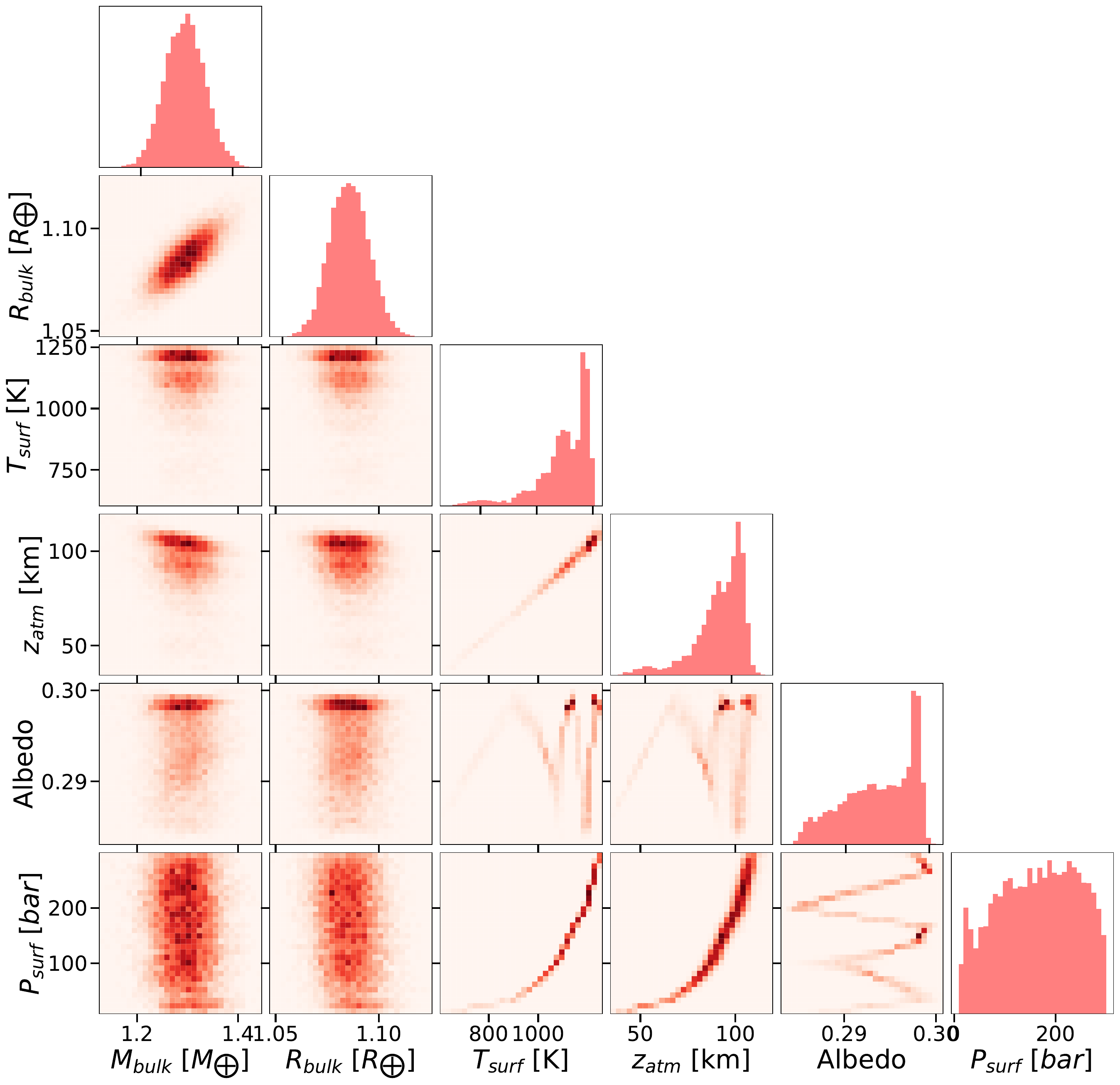}
                         }
      \caption{PDF corner plot of the atmospheric parameters of TRAPPIST-1 c for two different atmospheric compositions. The atmospheric parameters include the surface temperature ($T_{surf}$), atmospheric thickness ($z_{atm}$), Bond albedo, and the surface pressure ($P_{surf}$). The two envelope compositions we consider are H$_{2}$O (left panel) and CO$_{2}$ (right panel).}
      \label{fig:atm_trapp1c}
  \end{center}
\end{figure*}

\begin{figure*}[h]
        \centering
        \includegraphics[width=\hsize]{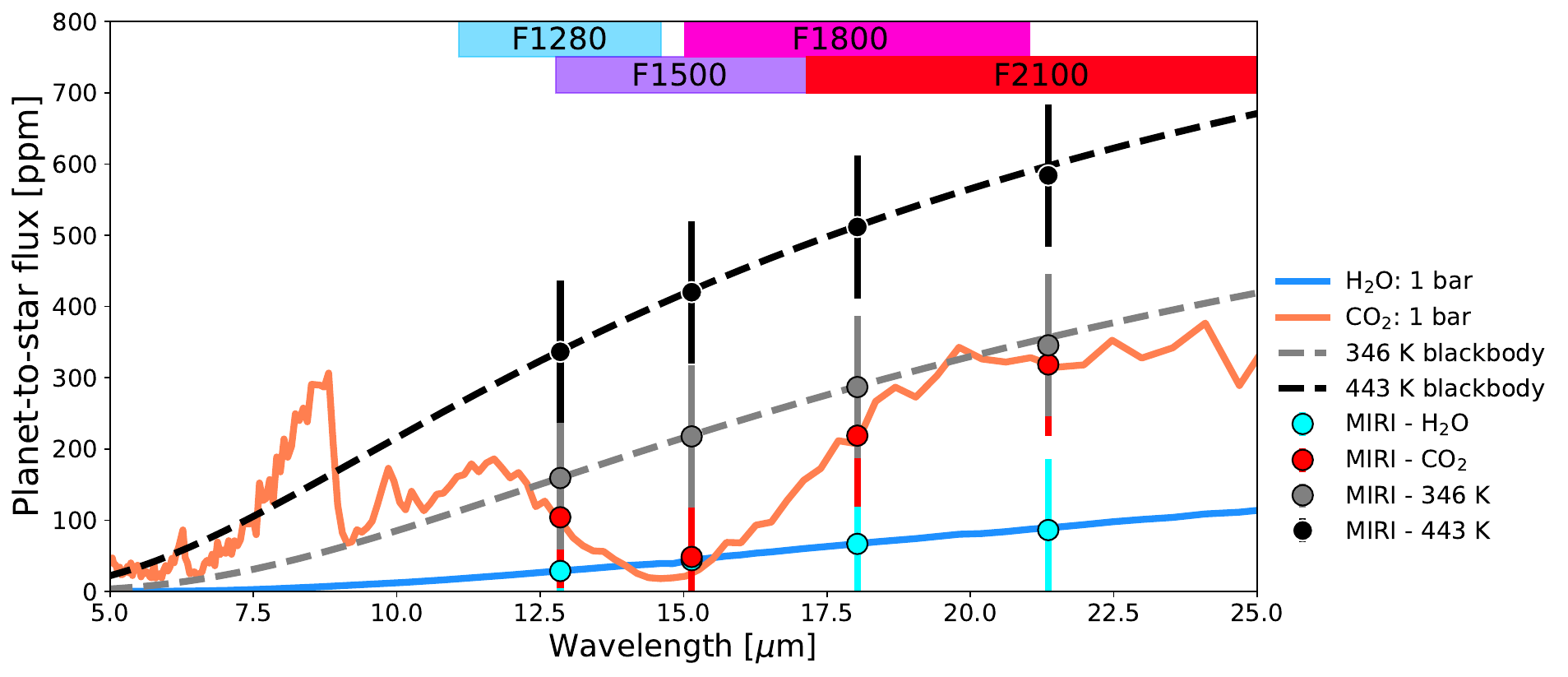}
        \caption{Simulated emission spectra of TRAPPIST-1 c with MIRI photometric filter mean fluxes for water and CO$_{2}$ atmospheres. The spectra were obtained with the high-resolution, k-correlated MSEIRADTRAN model. We show spectra for pure (100\%) water and CO$_{2}$ compositions. Colour boxes indicate the wavelength ranges of the different filters.}
        \label{fig:TRAP1c_phot}
\end{figure*}

TRAPPIST-1 c has been proposed to be observed in thermal emission with MIRI in JWST Cycle 1 \citep{Kreidberg_JWST21}. It will be observed with the F1500W filter during four eclipses, which is the filter centred at $\lambda$ = 15 $\mu$m. We analysed TRAPPIST-1 c with our adaptive MCMC for a water- and a CO$_{2}$-dominated atmosphere. Figure \ref{fig:atm_trapp1c} shows the marginal posterior distributions in 1D and 2D for this analysis. The CMF agrees well with the value obtained previously with our non-adaptive MCMC in \cite{Acuna21} and with the possible CMFs considered in \cite{Agol21}. \cite{Agol21} constrain the WMF with their MCMC and interior-atmosphere model with an upper limit of WMF = $10^{-3}$. This upper limit is two orders of magnitude larger than the mean of our PDF, showing a clear improvement in the resolution of the MCMC in the low surface pressure region of the parameter space. In their case, a maximum WMF = $10^{-5}$ estimate can only be shown with theoretical forward models (see their figure 18). In comparison with our own previous work, the distribution of the WMF derived by the adaptive MCMC is wider than the non-adaptive one, with mean and standard deviation values of WMF$_{adap}$ = 9.1$^{+4.4}_{-9.1}\times \ 10^{-5}$ when we do not consider the stellar Fe/Si constraint, and WMF$_{adap}$ = 3.3$\pm 3.3 \times \ 10^{-5}$ for the stellar Fe/Si scenario, while the non-adaptive value is WMF$_{non-adap}$ = (0.0$^{+2.7}_{-0.0}$)$\times \ 10^{-6}$. This difference is because the adaptive MCMC presents a higher acceptance rate due to a more effective exploration of the parameter space at WMF close to zero. More models in this region were explored, and thus more models are accepted in the posterior distribution, which becomes wider with a larger standard deviation than the non-adaptive posterior distribution. The corresponding surface pressure derived by the adaptive MCMC is $P_{surf}$ = 40$\pm$40 bar for a water-dominated envelope (Fig. \ref{fig:atm_trapp1c}). 

We can conclude that TRAPPIST-1 c could have an H$_{2}$O atmosphere of up to $\simeq$ 80 bar of surface pressure, an atmosphere with a less volatile species (CO$_{2}$, O$_{2}$), or no atmosphere at all. The analyses performed with the k-correlated atmospheric model indicate that an H$_{2}$O atmosphere in TRAPPIST-1 c would have a surface temperature between 1000 and 2000 K and an atmospheric thickness of 150 to 250 km. The posterior distributions of the surface temperature and atmospheric thickness show a bimodal shape in Fig. \ref{fig:atm_trapp1c} due to the inclusion of the stellar Fe/Si constraint in our retrieval. The peak at low values of $z_{atm}$ corresponds to low CMFs ($\simeq 0.20$), while the second peak corresponds to CMF ($\simeq 0.40$). If we only use the mass and radius as observable constraints in our retrieval, the first peak dissapears and the PDF becomes Gaussian.

For a CO$_{2}$-dominated atmosphere, we retrieved a CMF similar to the water case. For the volatile mass fraction, the PDF of the surface pressure is approximately uniform (see Fig. \ref{fig:atm_trapp1c}, right panel). The molecular weight of CO$_{2}$ is higher than that of water vapour, producing a more compressed atmosphere for a similar surface pressure and temperature. In addition, the radiative properties (i.e. opacity) of CO$_{2}$ yields a lower surface temperature for the same irradiation conditions in comparison to a water-dominated envelope, which contributes to a lower atmospheric thickness. As a consequence, the models with a CO$_{2}$ envelope can accommodate a more massive atmosphere for TRAPPIST-1 c than the water models, making it not possible to constrain the surface pressure of a CO$_{2}$-dominated atmosphere from mass and radius (and stellar Fe/Si) alone. We run a retrieval analysis with a different sampler, emcee \citep{emcee}, with a log-uniform prior for the surface pressure, and obtained a similar PDF for the CO$_{2}$-rich atmosphere.

We assumed the atmospheric parameters retrieved in our adaptive MCMC analysis and generated emission spectra with their respective temperature-pressure profiles, as explained in Sect. \ref{sec:obs_data}. Figure \ref{fig:TRAP1c_phot} shows the complete emission spectra and mean filter fluxes for TRAPPIST-1 c. Both our interior structure retrievals and evolution models \citep{KrissansenTotton21} predict that TRAPPIST-1 c is very likely to have a bare surface. Therefore, we also considered the possibility of a bare surface in TRAPPIST-1 c, given the high probability obtained in our MCMC analysis for a volatile mass fraction equal to zero. \cite{Hu12} obtained the emission spectra of bare terrestrial surfaces for different minerals. We estimated the brightness temperature for the irradiation conditions of TRAPPIST-1 c from the results of \cite{Hu12} for two minerals. These minerals are a metal-rich surface and a granitoid one since these are the two surfaces with the highest and lowest emission for the same irradiation conditions, respectively. We approximated the emission spectrum of these surfaces to that of a black body with a temperature equal to the estimated brightness temperature. These two brightness temperatures are 443 K and 346 K (black and grey in Fig. \ref{fig:TRAP1c_phot}, respectively). For very low emission fluxes ($<$ 200 ppm), TRAPPIST-1 c would present a CO$_{2}$-dominated atmosphere. For fluxes greater than 550 ppm in the F1500 filter, TRAPPIST-1 c would have no atmosphere and an emission that corresponds to the bare surface with the lowest albedo.

\cite{Zieba23} find that the emission of TRAPPIST-1 c in the F1500 filter is 421 $\pm$ 94 ppm. This value is within the range between 200 and 500 ppm, which presents a degeneracy between bare surfaces with high albedos, such as granitoid, feldspathic, and clay \citep{Hu12}, and thin atmospheres with little or no CO$_{2}$. Our water model shows an emission flux slightly above the CO$_{2}$ model in the 15 $\mu$m band. This may be due to the presence of clouds, which is considered in our 1D atmospheric model. Nonetheless, self-consistent 3D GCM models of water-dominated envelopes with surface pressures consistent with our 1$\sigma$ estimate show a higher emission flux \citep[][Turbet et al. In prep]{Lincowski_sub}. In this case, O$_{2}$ and H$_{2}$O-dominated atmospheres could be compatible with the measured emission of TRAPPIST-1 c, presenting a degeneracy between these, and a bare surface with high albedo. To break this degeneracy, follow-up photometric observations are needed at longer wavelengths (i.e filters F1800 and F2100) to distinguish between a bare rock, and O$_{2}$ and H$_{2}$O atmospheres. We observed that for the nIR filters ($\lambda = 5.60$ to $11.30 \ \mu$m), the models have very similar fluxes that are compatible within uncertainties, which makes distinguishing between the different scenarios in these wavelengths not possible.

\cite{Greene23} and \cite{T1b_emission} found that the emission flux of TRAPPIST-1 b is mostly consistent with a bare surface. This has implications for the Fe bulk content of all planets in this planetary system. In our analysis of TRAPPIST-1 c without a constrain on the Fe/Si mole ratio, we obtain that its CMF = 0.26 $\pm$ 0.08. This value is in agreement within uncertainties with the CMF derived for both planets b and c using the previous version of our modelling framework \citep{Acuna21}. The uncertainties in this work are larger to the previous analysis due to the adaptive sampling in the MCMC (see Sect. \ref{sec:mcmc}). This CMF constraint from planet b breaks the degeneracy between water and Fe content for the outer planets of TRAPPIST-1, supporting our WMF results for the outer planets of the system in scenario 2 of \cite{Acuna21}.

\subsection{55 Cancri e}




\begin{table*}[h]
\caption{MCMC retrieved mean value and 1$\sigma$ uncertainties of the observable parameters (data), compositional parameters (core and volatile mass fractions), and atmospheric parameters (boundary interface temperature, and atmospheric thickness) of 55 Cancri e.}

\centering
\begin{tabular}{ccccc}
\hline
                           & Data          & With stellar Fe/Si  & Without stellar Fe/Si  & No volatile layer    \\ \hline
\multicolumn{1}{c|}{$M$ [$M_{\oplus}$]}     & 7.99$^{+0.32}_{-0.33}$  & 7.84$^{+0.32}_{-0.33}$  & 7.88$\pm$0.33 & 8.12$\pm$0.29 \\
\multicolumn{1}{c|}{$R$ [$R_{\oplus}$]}     & 1.875$\pm$0.029 & 1.898$\pm$0.022 & 1.892$\pm$0.029 & 1.860$\pm$0.021 \\
\multicolumn{1}{c|}{Fe/Si} & 0.60$\pm$0.14 & 0.66$\pm$0.12 & 1.41$\pm$0.50 & 0.12$^{+0.19}_{-0.12}$\\
\multicolumn{1}{c|}{CMF}   &               & 0.22$\pm$0.04 & 0.38$^{+0.09}_{-0.10}$ & 0.06$^{+0.07}_{-0.04}$\\
\multicolumn{1}{c|}{WMF}   &               & $(5.0^{+1.6}_{-5.0}) \times 10^{-3}$ & 0.010$^{+0.008}_{-0.010}$ & 0.0 (constant)\\
\multicolumn{1}{c|}{ $T_{300 \ bar}$ [K]}    &               & 4427$\pm$3 & 4422$\pm$3   \\
\multicolumn{1}{c|}{$z_{atm}$ [km]}   &               & 606$\pm$20 & 550$^{+34}_{-19}$   \\ \hline
\end{tabular}
\label{tab:55cnce_res}
\end{table*}


The super-Earth 55 Cancri e is also in close orbit (P = 0.66 days) to a bright star. This exoplanet has had several interior and atmospheric hypotheses proposed. \cite{Madhusudhan12} explored a carbon-rich interior given the high C/O ratio found for the host star, showing that in this case, the planetary bulk density would be lower than that of a silicate-rich mantle planet, such as Earth. They concluded that a volatile layer would not be necessary to account for its density. However, a classical Fe-rich core and a silicate mantle are compatible with a volatile envelope rich in secondary atmosphere species. Furthermore, the absence of an H/He-dominated envelope seems likely to be due to the lack of hydrogen and helium emission and absorption lines in the spectrum \citep{Ehrenreich12,Zhang21}. The presence of a secondary atmosphere is supported by phase curve data from the \textit{Spitzer} Space Telescope \citep{demory11,Angelo17} and 3D GCM modelling. The latter suggests that 55 Cancri e could have an optically thick atmosphere with a low mean molecular weight \citep{Hammond17}. The possibility of a fully H$_{2}$O-dominated atmosphere was discarded since it would require the presence of water and hydrogen simultaneously in the atmosphere due to water dissociation. Possible compositions for the atmosphere of 55 Cancri e are a mixture of silicate compounds \citep{Keles22}, such as HCN, detected by \cite{Tsiaras16}, with traces of water \citep[detected by][]{Esteves17} or CO$_{2}$, CO, and N$_{2}$, among other compounds.


\cite{Hu_jwstprop_cnc} have proposed to observe 55 Cancri e in emission spectroscopy combining the NIRCam F444W filter (3-5 $\mu$m) and MIRI LRS (5-14 $\mu$m). We summarize the 1$\sigma$ confidence intervals of the interior and atmospheric parameters obtained in our retrievals for 55 Cancri e in Table \ref{tab:55cnce_res}. We observed that a water-dominated atmosphere reproduces the observed data well, with WMF up to 1.8\%, indicating the possibility of a thick envelope with $P_{surf} > 300$ bar.

We do not model the scenario of a CO$_{2}$-dominated envelope for 55 Cancri e. The reason is twofold: 1) CO$_{2}$ envelope is not extended enough to match the density of 55 Cancri e, yielding a denser interior than the data show, and 2) at temperatures higher than 4000 K, CO$_{2}$ would not be the dominant species in a C-rich atmosphere, but CO. This changes the emission of the atmosphere, as CO is a different absorber from CO$_{2}$. A CO-rich atmosphere could also explain the low density of 55 Cancri e in this scenario since CO has a lower molecular weight than CO$_{2}$ and would yield a larger atmospheric scale height. When H/He is not included in the interior modelling,  water as a trace species is necessary to explain the low density of 55 Cancri e since a purely dry silicate atmosphere would have a smaller thickness than a CO$_{2}$ atmosphere due to heavier molecular weights under similar atmospheric surface conditions. Adding silicate absorbers decreases the total planetary radius in H/He envelopes \citep{Misener21}. However, more modelling work is necessary to explore the effect of silicates in atmospheres that have lost their primordial H/He. 

A planet with no volatiles matches the low planetary density for CMFs below 13\% (Table \ref{tab:55cnce_res}), which is indicative of a bulk interior less dense and Fe-rich than that of Earth ($CMF_{\oplus} = 32\%$). An alternative bulk composition for 55 Cancri e in the volatile-poor scenario would be a carbon-rich mantle, as suggested by \cite{Madhusudhan12}.

\begin{figure}[h]
        \centering
        \includegraphics[width=\hsize]{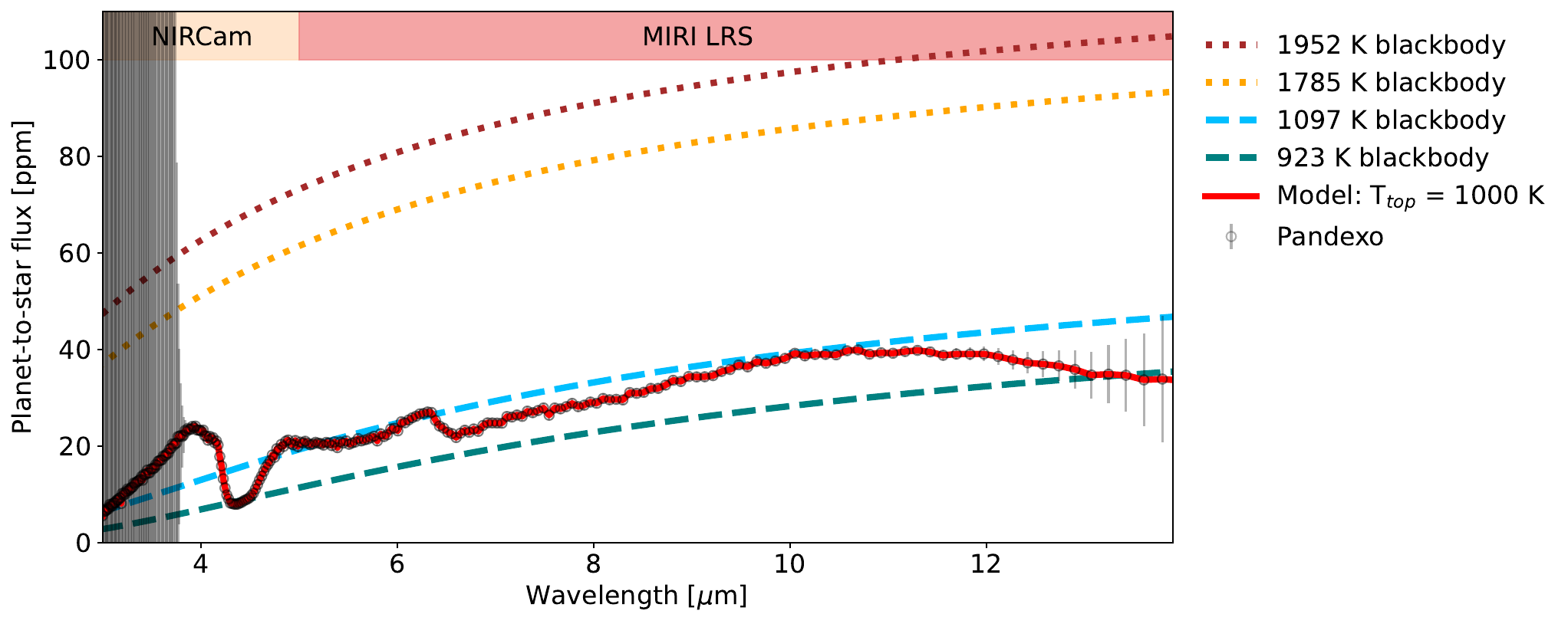}
        \caption{Predicted emission spectrum with the k-correlated, high-resolution MSEIRADTRAN model for a water-rich atmosphere in 55 Cancri e with NIRCam and MIRI LRS. We assumed a mesosphere temperature $T_{top}$ = 1000 K. We show the black body emission at different irradiation temperatures for comparison. 
        }
        \label{fig:55cnce_spectrum}
\end{figure}

In most of the applications of our modelling framework, we consider a mesospheric temperature of $T_{top} = 200$ K. 55 Cancri e is extremely irradiated, and it is likely to have a higher mesospheric temperature. We repeat our analysis with a mesospheric temperature of $T_{top} = 1000$ K. We find no difference in the interior parameters compared to the low $T_{top}$ analysis. We consider the high top temperature case to generate the emission spectrum at higher resolution. Fig. \ref{fig:55cnce_spectrum} shows the complete predicted emission spectrum of 55 Cancri e from 3 to 14 $\mu$m. 
The water line at 4.3-4.4 $\mu$m can be seen. At wavelengths above 3.7 $\mu$m the noise level is low (Fig. \ref{fig:55cnce_spectrum}), which makes the spectral features in this range easy to identify with JWST in the high-molecular weight atmosphere scenario of 55 Cancri e. For comparison, in Fig. \ref{fig:55cnce_spectrum} we also show the blackbody emission at 1952 and 1785 K, which correspond to the irradiation temperautures typically associated to bare rocks \citep{Hu12}.

\section{Discussion}
\label{sec:discussion}

In this section, we discuss the processes that were not included in our model but that may affect our estimates on the volatile mass fraction, such as solubility in a magma ocean and near-surface isothermal layers. TRAPPIST-1 c and 55 Cancri e are warm enough to hold a magma ocean on their surface. The dissolution of silicates and the density of the mantle can change between a dry mantle model and that of a wet magma ocean \citep{DL21}. For a planet less massive than TRAPPIST-1 c (0.8 $M_{\oplus}$) and a radius of $\simeq 1.1 \ R_{\oplus}$, a dry mantle model and a wet magma ocean model estimate a WMF = 3 $\times 10^{-3}$ and 8 $\times 10^{-3}$, respectively. This difference decreases with a greater planet mass and a lower WMF. Including dissolution of silicates and a magma ocean would increase our WMF estimates by a factor of less than  2.7 \citep[see figure 4 in][]{DL21}. Since our MCMC realisations span at least two orders of magnitude (from $10^{-6}$ to $10^{-4}$), the wet magma ocean model would only shift our WMF distribution by less than one order of magnitude. The maximum surface pressure of TRAPPIST-1 c would go from 25 bar to 75 bar, at most. A retrieval with a wet magma ocean would not discard a bare surface in TRAPPIST-1 c since the radius of a wet melt-solid interior is less than that of a dry solid interior, leaving more room for an atmosphere above. For 55 Cancri e, which has a mass of $8 \ M_{\oplus}$, the difference is negligible, and it would yield similar WMF estimates if we considered a wet magma ocean surface.

In our atmospheric model, we prescribe the atmosphere's thermal structure as a near-surface, dry convective layer followed by a wet convective layer and an isothermal mesosphere. A self-consistent treatment of the shortwave radiation together with an iterative scheme on the temperature profile would enable us to compute the regions of the atmosphere where radiative layers would form as well as  their exact temperature values. We tested how changes in the mesospheric temperature may impact the OLR and the atmospheric thickness with our atmospheric model. We performed a test with a new mesospheric temperature of 1000 K, and compare it to our default temperature of 200 K. The surface temperature at which the atmosphere is in radiative-convective equilibrium is similar in the two cases, yielding similar atmospheric thicknesses. Therefore, our envelope mass fraction estimates are robust against different upper radiative temperatures. In contrast, near-surface radiative layers may decrease the thickness of the atmosphere, compared to a convective atmosphere. However, such layers are more likely to form in atmospheres composed of H/He and silicates than pure water or CO$_{2}$ envelopes \citep{misener22}. This means that our envelope mass fraction estimates are lower limits compared to those that would be obtained with an atmosphere that presents near-surface radiative layers. Moreover, \cite{vazan22} find that water envelopes with silicates only develop radiative layers at low pressures (100-10 bar; see their figure 7) and not at the near-surface. 

In our interior analysis, we consider a grey model for the calculation of the albedo (Sect. \ref{sec:albedo}). The estimated grey opacity for water is based on Earth \citep{Nakajima92}. Nonetheless, this parameter may be different for highly-irradiated planets. Our interior-atmosphere models yield albedos between 0 and 0.30 in the water envelope case for TRAPPIST-1 c and 55 Cancri e. The change in surface temperature induced by this variation in the albedo is below 30 K. This difference in surface temperature produces changes in radius of less than 1\% (Sect. \ref{sec:validation}). Planets with a magma ocean underneath their envelope may present hazes and aerosols \citep{Kempton23}, whose albedo is higher than that of Earth-like water clouds. The effects of different cloud properties on interior modelling has been explored for gas giants by \cite{Poser23}, who found that if optically thick clouds are high up in the upper atmosphere, they have a negligible effect in the inference of metal (water, rock) content.

To calculate the density of CO$_{2}$, we used the ideal gas EOS, in contrast to a non-ideal EOS for water. In the following, we discuss how the use of an ideal EOS for carbon dioxide may affect our results. \cite{bottcher12} carried out a comparison between the ideal EOS and non-ideal EOS for carbon dioxide. They find that CO$_{2}$ starts to behave as a non-ideal gas at $\simeq$ 7 MPa, which corresponds to 70 bar.  The non-ideal EOS yields a higher density than the ideal EOS \citep{bottcher12}. This means that the atmospheric thickness of a CO$_{2}$ envelope with a non-ideal EOS would be even lower than that calculated with the ideal EOS. This would make it even harder to match the current density of 55 Cancri e with a CO$_{2}$-dominated atmosphere,  which further supports that CO is more likely to constitute the atmosphere of this planet instead. For TRAPPIST-1 c, a higher density of the CO$_{2}$ envelope in comparison to the water envelope strengthens the degeneracy with the bare rock scenario. The implementation of a non-ideal EOS, such as the SESAME EOS 5210 for carbon dioxide \citep{sesame}, and the \cite{coeos} EOS for CO, will be the focus of future work.

\section{Conclusions}
\label{sec:conclusion}



In this work, we present a self-consistent model built to estimate the internal compositions and structures of low-mass planets with water and CO$_{2}$ atmospheres when given their observed mass, radius, and their host stellar abundances. We coupled the interior and the atmosphere self-consistently to obtain the boundary conditions at the top of a supercritical water layer or a silicate mantle given the irradiation conditions of a low-mass planet. This was done by calculating the bolometric emission flux and the Bond albedo in order to compute the flux emitted and absorbed by an atmosphere in radiative-convective equilibrium. We used a 1D k-correlated atmospheric model with updated opacity and EOS data for the computation of the bolometric emission within our interior-atmospheric model. We also demonstrated that using a constant step size when sampling the prior distribution in a MCMC scheme is not efficient for exploring the parameter space in interior modelling. This constant maximum step size causes an underestimation of the uncertainties of the compositional parameters. Therefore, it is necessary to use an adaptive MCMC when performing retrieval with interior models, especially for planets whose compositional parameters can reach the maximum or minimum possible values. This is the case for rocky Earth-sized planets and super-Earths, whose WMFs are close to zero but nonetheless remain important for determining surface pressure.

Moreover, we used the surface pressure and temperature conditions retrieved with our interior-atmosphere model to generate emission spectra with our k-correlated atmospheric model, MSEIRADTRAN. We computed emission spectra to show how the output of our interior-atmosphere model can be used to predict the input necessary for atmosphere models and help prepare atmospheric characterisation proposals. The particular parameter that interior models can provide self-consistently for atmosphere models are the surface pressure and temperature, which are usually chosen arbitrarily in order to generate spectra. Emission spectra are more sensitive to the choice of surface temperature and thermal structure than transmission spectra.

We showcased how to use interior and atmospheric modelling simultaneously to predict observations for two rocky planets, TRAPPIST-1 c and 55 Cancri e, which have been proposed for emission photometry and spectroscopy observations with JWST. We binned our emission spectra according to the response functions of the MIRI filters in order to predict emission fluxes for TRAPPIST-1 c in different scenarios, while for 55 Cancri e, we input our emission spectra to Pandexo to predict observational uncertainties. 


The most likely scenario for TRAPPIST-1 c is that it lacks an atmosphere (WMF = 0). Nonetheless, the presence of a secondary atmosphere cannot be ruled out. In this scenario, TRAPPIST-1 c could have an H$_{2}$O-dominated atmosphere of up to 80 bar of surface pressure. The density of a CO$_{2}$ or O$_{2}$-rich envelope is not low enough to put any constraints on the surface pressure from mass and radius data alone. We presented emission flux estimates for the filter centred at 15 $\mu$m, F1500, that can be compared with observations \citep{Zieba23}. Moreover, we discuss that a bare surface in TRAPPIST-1 b \citep{Greene23,T1b_emission} has implications for the Fe content of all planets in the system. The Fe/Si mole ratios for planets b and c as retrieved in \cite{Acuna21} and revisited in this work can be used to constrain the Fe contents of the other planets in the TRAPPIST-1 system. This breaks the degeneracy between WMF and CMF for the outer planets in TRAPPIST-1, supporting the WMFs obtained in scenario 2 of \cite{Acuna21}. For 55 Cancri e, a massive water envelope with more than 300 bar of surface pressure is necessary to fit its low density and a Fe content similar to its stellar host or an Earth-like core simultaneously. We determined that a combined spectrum with NIRCam and MIRI LRS, as proposed by \cite{Hu_jwstprop_cnc}, may present a high noise level at wavelengths between 3 and 3.7 $\mu$m. However, this part of the spectrum does not contain any spectral lines of water or CO$_{2}$, which are essential to determine the abundances in the envelope.

In our modelling approach, we have considered water- and CO$_{2}$-dominated atmospheres (99\% H$_{2}$O and 1\% CO$_{2}$ and vice versa). However, the atmospheres of low-mass planets are more diverse than these two compositional scenarios. The atmospheric compositions of sub-Neptunes are proving to be a mixture of H/He, water, and other compounds, according to observations and models \citep{Madhusudhan20,Bezard20,guzmanmesa22}, while super-Earths can have more exotic atmospheric compositions, such as mineral atmospheres \citep{Keles22}. Therefore, the aim of future work will be to include more gases in the atmospheric model as well as the calculation of transmission spectra in addition to the existing implementation of emission and reflection spectra. Our interior-atmosphere model, MSEI, serves as a precedent to develop models with more diverse envelope compositions in order to prepare proposals for JWST and future atmospheric characterisation facilities, such as Ariel \citep{Tinetti18}. Our model can also be used within retrieval frameworks to simultaneously interpret mass, radius, and upcoming JWST emission spectral data to break degeneracies in exoplanet compositions.

\begin{acknowledgements}
M.D. and O.M. acknowledge support from CNES. We thank the anonymous referee whose comments helped improve and clarify this manuscript. We acknowledge Emmanuel Marcq and Jérémy Leconte for their exchange on the k-correlated method. L. A. thanks Paul Mollière and Laura Kreidberg for useful discussions on emission spectra.
\end{acknowledgements}

\bibliographystyle{aa} 
\bibliography{biblio}  

\end{document}